\documentclass[a4paper,12pt]{article}
\usepackage[utf8]{inputenc}
\usepackage{cancel}
\usepackage{tikz}
\usepackage{ulem}
\usepackage{amsfonts}
\usepackage{amssymb}
\usepackage{graphicx}
\usepackage{amsmath}
\usepackage{enumerate}
\usepackage{mathtools}
\usepackage{subfig}
\usepackage{color}
\usepackage{tikz}
\usepackage{float}
\usepackage{here}
\usepackage{cite}
\usepackage{mathrsfs}
\usepackage{float,epsfig}
\usepackage{dcolumn}
\usepackage{graphicx}
\usepackage{bm}
\usepackage{amsmath,amssymb,amsthm}
\usepackage[colorlinks=true,linkcolor=blue,citecolor=red]{hyperref}
\usepackage{multirow}
\usepackage[toc,page]{appendix}
\usetikzlibrary{arrows.meta}
\usetikzlibrary{bending}
\usetikzlibrary{calc}
\newcommand{\be}{\begin{equation}}
\newcommand{\ee}{\end{equation}}
\newcommand{\bea}{\setlength\arraycolsep{2pt} \begin{eqnarray}}
\newcommand{\eea}{\end{eqnarray}}

\def\0{{\sst{(0)}}}
\def\1{{\sst{(1)}}}
\def\2{{\sst{(2)}}}
\def\3{{\sst{(3)}}}
\def\4{{\sst{(4)}}}
\def\5{{\sst{(5)}}}
\def\6{{\sst{(6)}}}
\def\7{{\sst{(7)}}}
\def\8{{\sst{(8)}}}
\def\sst#1{{\scriptscriptstyle #1}}

\makeatletter \@addtoreset{equation}{section}

\definecolor{lime}{HTML}{A6CE39}

\begin{document}

\title{{\normalsize \textbf{\Large Swampland Statistics for Black Holes  }}}
\author{ {\small  Saad Eddine Baddisi\thanks{Corresponding author: saadeddine.baddis@um5r.ac.ma} ,    Adil  Belhaj, Hajar Belmahi\thanks{
 Authors are listed  in alphabetical order.} \hspace*{-8pt}} \\
%EndAName
{ \small Faculty of Science, Mohammed V University in Rabat, Rabat,
Morocco}}
\maketitle

\begin{abstract}
 In this work, we  approach certain black hole issues, including remnants, by  providing  a statistical description   based on  the weak gravity conjecture in the swampland program. Inspired by  the  Pauli exclusion principle in the context of the Fermi sphere, we derive an inequality which  can be exploited  to verify the instability manifestation of non-supersymmetric four dimensional black holes via a characteristic function. For several species, we show that this function matches with   the weak gravity swampland conjecture. Then,   we  deal with the cutoff issue  as an interval estimation problem by putting a lower bound  on the black hole mass scale  matching with certain results reported  in the  literature. Using the developed formalism for the proposed instability scenarios, we present a suppression mechanism to the remnant production rate. Furthermore, we reconsider  the stability study  of the Reissner–Nordström black holes.  As a result,  we show that the approached   instabilities prohibit naked singularity  behaviors. \\
\textbf{Keywords}: Black holes, Swampland conjectures, Fermi sphere, Hawking evaporation, Species scale.
\end{abstract}

\newpage

\section{Introduction}
 Recently, a special interest has been devoted to the study of effective field theories (EFTs) from higher dimensional supergravity  scenarios  such as   superstrings and  M-theory inspired models in connection with the swampland program \cite{INTM2,INTM6,INTM7,INTM8,INTM9,INTM10,INTM11,INTM12,INTM13,INTM14}. The latter aims to address such  EFTs through discussions on the superstring landscape. Moreover,   it  could  be used to check for consistencies  with quantum gravity (QG).  Moreover, this program could  provide an alternative falsification mechanism  in  the light of  the absence of  the empirical evidence for certain  physical theories. A close inspection shows that many  swampland criteria   have been suggested, developed, and refined  in the context  of superstring compactifications. Generally, these compactifications  generate   various  scalar fields in the associated spectrum. Some of them   correspond  to appropriate  geometric concepts  including the size and the shape of   internal compact  spaces with non-trivial holonomy groups as well as   the Calabi-Yau manifolds  \cite{INTM15,INTM16, INTM17,INTM18}.    These scenarios have been largely elaborated in order to check  and  verify  the application of the suggested swampland conditions. In  contact  with such developments, two  relevant conjectures  have been investigated being known by distance  conjecture (DC) and weak gravity conjecture (WGC). The first one concerns the implication of  the  stringy moduli  space in the  EFT building models. In superstring theory,  for instance, this space is parametrized by massless scalar fields originated from the    compactification mechanism dealing with the  dimension reduction of the space-time in which   the  supergravity spectrums live.  This   physical road  could predict certain light particles and towers at large moduli space distances   according to T-duality arguments and stringy  size  behaviors \cite{INTDC4,INTDC5,INTDC6,INTDC7,INTDC8,INTDC9,INTDC10,INTDC11,INTDC12,INTDC13}.  The second conjecture (WGC) was   first  proposed in \cite{INTWGC1}. After that, it has been developed  and approached from different angles\cite{INTWGC2,INTWGC3,INTWGC5,INTWGC6,INTWGC7,INTWGC8,INTWGC9}. This conjecture stems from the fact that gravity is the weakest fundamental interaction which is translated into two main observations. First, a cutoff scale at which the low energy EFT breaks is bound from the top by a gauge coupling. Secondly, the stable black hole horizons must have little to no charge compared to black ole mass. Otherwise at extremal limits the horizon become unstable, effectively suppressing naked singularities. Moreover, the correspondence between the WGC and the DC had been explained in \cite{DCWGC1,DCWGC2,R11,R12}. Specifically, the saturation of WGC bounds by extending the moduli distance is a common observation.

In black hole physics in arbitrary dimensions,  WGC has been manipulated  to put  certain constraints on the charge per unit  mass \cite{R11,R12,R13}.  It  has been also exploited to investigate  small black  holes in four dimensions derived from string theory using brane physics \cite{BRANE1}.   Moreover,   the primordial black holes  and the  black hole species  have been  also approached in connection with   WGC constraints. In particular,    it has been  revealed that  the primordial  extremal black holes  with appropriate   mass conditions   could be considered as dark matter candidates \cite{PREBLACK1,PREBLACK2}. Beside that, the black holes evaporating down to the Planck scales have been studied in the light of the UV cutoff problems and the no-global symmetry conjecture \cite{INTWGC1}.

The aim of this work is to contribute to these activities by approaching certain black hole issues, including remnants, by  providing  a statistical description relaying on  WGC  in the swampland program. Inspired by  the  Pauli exclusion principle in the context of the Fermi sphere, we derive an inequality  which can be utilized  to verify the instability manifestation of the black holes by help of a statistical characteristic function. For  a generic number of species, we show that this function is in agreement with  the weak gravity swampland conjecture. Next, we examine the possibility of using the cutoff as an interval estimation problem by imposing an upper limit on the black hole mass scale, which is compatible with some results obtained in the literature. \cite{SPE1,SPE2,SPE3,SPE4,SPE5,SPE6,SPE7,SPE8}.
 Using the developed formalism for the proposed instability scenarios, we provide  a remnant production rate suppression mechanism. Moreover,  we reconsider  the stability behaviors of the Reissner–Nordström black holes. Precisely, we show that the proposed instabilities prohibit naked singularity  behaviors.\\

The present  work is  structured as follows. In section 2,  we  present a concise discussion of black hole remnants.  In section 3, we attempt to derive  quantities  allowing one  to  elaborate certain  aspects of  WGC and approach the remnant problems.   In section 4, we   present  a  swampland statistical description  for  the  Reissner–Nordström black holes. The last section summarizes the main conclusion and provides some   open questions.

\section{Remnant production rate}
In this section, we would like to address  issues associated with the contemporary theory of black holes. In particular, we discuss  the corresponding remnant problems.  As the name hints, the remnants are  evaporating black hole remains.  They  have been proposed to   overcome  the information loss problem in black hole physics.    Various attempts have been made to solve this problem. Concretely, a  black hole formation description was    first proposed in \cite{Rem1}  by establishing   the expression form  $\frac{M_p^3}{\pi^3}e^{-\frac{M_p^2}{16\pi T^2}}$, where  $M_p$ is the Planck mass. Later,  this work has been  developed by providing a  classical description of such a formation\cite{Rem2}. In these contributions,   the formation rate of the  black holes has  been derived, at a finite temperature  being related to the black hole mass via the Hawking radiation, $T=\frac{M_p^2}{8\pi M}$.  According to  the works developed in \cite{Rem3}, these scenarios  have been linked to the remnant production rate via the   following form 
\begin{equation}
P\sim N_Re^{-\frac{M^2}{M_p^2}}
\end{equation}
where $N_R$ is the number of different remnant species. $M$ is the mass  of  the black hole. The main issue of this description  is that the statistics of such objects do not add up. In fact,  it has been observed that the production  probability  is very large  for remnants at small mass scales including the Planck one. In any case, the number of remaining species can be infinite, as there is no fixed limit for this number. In certain cases, it has been observed that the situation can be exacerbated when the internal entropy of the remains is used to count $N_R$. In order to avoid altering the Bekenstien-Hawking entropy, it has been proposed that the remaining population could take the following form 
\begin{equation}
N_R\sim e^{2r M_p},
\end{equation}
where $r$ is the distance of a fiducial observer in the thermal atmosphere \cite{Rem4}. In this  way, the remnants  are  taken to be in the Planck scale. However, lower mass values could be considered.  Additionally,    at such small scales,   the  distance  has been  identified with  the internal entropy of the black hole  $S_R$ through  the relation    $r=\frac{S_R}{M_p}$. Therefore, the remnant production rate   can be  governed by the  black hole internal entropy,   being  an increasing  quantity,  muddying the water further.\\

From another perspective, in \cite{Rem5}, it was pointed out that when the content of a black hole is separated into several quantum states, called “informons,” the decay rate of a weakly coupled electric field into these informons (the rate of residue production) includes a divergent term in the form of a sum of exponential expressions over the number of these states. This divergence would be caused by the infinite number of informons whose mass is less than the Planck scale.
As a prelude to what follows, this question will be addressed by proposing that it is not possible to fit an infinite number of states into the black hole. Alternatively, the usual Hawking evaporation would be interrupted by the decay of the black hole along certain probable decay channels.

In what follows, we discuss some considerations of black holes, including remnants, by offering a statistical description based on the WGC within the swampland program.\\

\section{WGC from the asymmetric instability}
\label{sec:2}
For letter use, we would like  to present, first,  a  statistical scenario  of the instability behaviors  inspired by WGC being   originally  suggested to solve  certain  black hole problems including the remnant  ones \cite{INTWGC1}. Roughly,  this conjecture stands on two pillars, motivating the present study. First, it addresses EFT cutoffs $ \Lambda$   by providing a QG obstruction to a vanishing value of the gauge coupling $g$, prohibiting the restoration of  the  abelian  phase  global symmetry \cite{R12}. This is ensured by implementing   the following four dimensional inequality
\begin{equation}
 \Lambda\leq gM_p.
 \label{WGC}
 \end{equation} 
 Secondly, it   could solve the  naked singularity problems by predicting the decay of the extremal black hole into EFT stable particles. \cite{INTWGC1}. Explicitly,  in  gauge theories  weakly coupled to gravity, there exist charged states of a mass $m$ and  an electric charge $q$  satisfying  the inequality
 \begin{equation}
 \frac{q}{m}\geq\frac{Q}{M}\vert_{\mbox{extermal}}=\mathcal{O}(1)
 \end{equation}
where  $Q$   is  the  black hole charge \cite{R12}.
 Instead of continuing to evaporate and consequently becoming a naked singularity, this inequality ensures that the extremal black hole decays to preserve regular horizons described by the identity
 \begin{equation}
  \frac{Q}{M}<\mathcal{O}(1).
  \end{equation} 
This  WGC motivation could be exploited to  address  the remnant problems by introducing an instability   analysis. This   concerns  the  black hole Hawking evaporating down to the Planck scale.
  
In the present investigation, we attempt  to approach the problem of remnants via  WGC using  a statistical  description.  To do so, we first reconsider the study of such a conjecture through a suggested asymmetric instability scenario.  This discussion will be based on    the  following proposed statement.  In a stable (non decaying) cavity, the number of the accessible charged states should not exceed the available number of massive states representing the entirety of the system. This statement  has been inspired by the  Pauli exclusion principle which has   been exploited to introduce the asymmetric instability (or asymmetric energy) in the semi-empirical mass model \cite{Book1},  where  the higher energy levels are occupied by less stable states. Motivated  by the  Fermi description of the energy of a ball of non relativistic $\frac{1}{2}$-spin fermions \cite{Book2}, we would like  to provide a possible  description of such instabilities. Concretely, this will be done using  a Fermi spherical description.  In this way,  the number  of charged states  can be expressed as 
\begin{equation}
z=\iint \frac{\prod\limits^2_{\alpha=1}dx_\alpha dp_\alpha}{\hbar^2},
\end{equation}
where $z$  can be identified with  the size parameter of the phase space described by the fibration  $S^2\times \mathbb{R}^2$.   It is also called the number density being a continuous parameter\cite{Book2}. It is worth nothing that $S^2$ corresponds to the black hole surface  and $\mathbb{R}^2$ is the momenta space of  a particle in such a surface.\\

In the following, we use  the expressions $dx_1dx_2=r^2\sin(\theta)d\theta d\phi=dA$ and $ dp_1dp_2=dp_\theta dp_\phi=d^2p$ with the condition $p_\theta=p_\phi=p$.  By deriving the number of states with respect to the momentum and integrating over the
spatial variables, we arrive at the following Fermi sphere law
\begin{equation}
\frac{dz}{dp}=\frac{Ap}{\hbar^2}.
\label{eq1}
\end{equation} 
This law basically dictate how one should count the states in the given phase space.\\
In the   black hole  physics,   for instance, the quantity $A$   can be identified  with  the event horizon area of a Schwarzschild black hole, given by  $A=\frac{16\pi G^2M^2}{c^4}$ \cite{Book4}.   Since we wish to establish a scenario of instability governed by a charge-to-mass ratio, we will not delve into more detailed phenomenological studies of these states. We will therefore consider that the observer counting the states is in a reference frame where relativistic effects are negligible. Combining Eq.(\ref{eq1}), and the relation $p=\sqrt{2m_gE_z}$, the energy associated with the asymmetric instability is   found to be 
\begin{equation}
E=\int E_zdz=\frac{\hbar^2c^4z^2}{32\pi G^2M^2m_g},
\end{equation}
where $m_g$ is the mass of a unit charge (particle/specie)  of a gauge  coupling $g$.  This quantity  is needed to define  the  characteristic function  being the  inverse of the production rate of the specie $(g,m_g)$ at a  thermodynamic equilibrium.  Indeed, it can be expressed as follows
\begin{equation}
 \rho=\frac{1}{\mathcal{Z}}e^{-\frac{\hbar^2c^4z^2}{32\pi k_{B}TG^2M^2m_g}}
 \label{eq2}
 \end{equation} 
where $T$ is the temperature and $\mathcal{Z}$ is  the   corresponding  partition function \cite{Book1}. The normalization constraint of the characteristic function leads to  
\begin{equation}
 \mathcal{Z}=\iint e^{-\frac{\hbar^2c^4z^2}{32\pi k_{B}TG^2M^2m_g}}\frac{r^2\sin(\theta)d\theta d\phi dp_\theta dp_\phi}{\hbar^2}.
\end{equation}
Using  integration techniques, we  obtain 
\begin{equation}
\mathcal{Z}=\frac{32\pi k_{B}T G^2M^2m_g}{z\hbar^2c^4}.
\end{equation}
The characteristic function given by Eq.(\ref{eq2}) can be treated as a probability of accessing a state $(M,g,m_g)$.  In fact,  the instability can be envisaged on the basis of decreasing values linked to the probability of accessing a flow of species produced in pairs on small scales.  In order to  examine such behaviors,  we  need to   calculate the decay rate. Indeed,  we get 
\begin{equation}
\frac{\partial \rho}{\partial(\frac{1}{M^2})}=\frac{zc^4\hbar^2}{36\pi k_BTG^2m_g}(1-\frac{c^4\hbar^2z^2}{36\pi k_BTG^2M^2m_g})e^{-\frac{\hbar^2c^4q^2}{32\pi k_{B}TG^2M^2m_g}}.
\end{equation}
The asymmetric instability corresponds to the constraint  $\frac{\partial \rho}{\partial(\frac{1}{M^2})}<0$. This is insured by the inequality
\begin{equation}
\frac{c^4\hbar^2z^2}{32\pi k_BTG^2M^2m_g}>1.
\end{equation}
Considering quantized total charge $Q=zg$, and using the Planck units and the  global charge, we have
\begin{equation}
\frac{Q^2}{M^2}>\frac{32\pi g^2m_g T}{ M^3_p T_p}.
\label{ineq1}
\end{equation}
Taking $M=NM_p$, the inequality corroborates the proposed statement up to a factor $\sqrt{\frac{32\pi m_g T}{M_p^2 T_p}}$. Indeed, the proposed instability stems from the fact that the number of charged states $z$ exceeds the available number of massive states $N$. However, one should  factor in the characteristics of the cavity, being the mass for the contained states $m_g$, the equilibrium temperature $T$, and the corresponding Planck scales.

Moreover, the instability of the  states satisfying the inequality (\ref{ineq1})  increases  when there are species  maximizing  the ratio $\frac{z}{m_g}$.  
 This   can be summarized by the following statement: A consistent EFT  should always have species that maximize the ratio $\frac{\vert q_i\vert}{m_{i}}$, with  $i=1, \ldots , N_{sp}$  where  $N_{sp}$ is the number of  the species  and one has $q_i=gz_i$. Moreover, one considers that the black hole is composed of non-relativistic fermions with different integer spins. Thus, we can  decompose the net charge $Q$ into elementary charges of species $N_{sp}$.  According to   \cite{R11},  this has been formulated   as follows
\begin{equation}
\frac{\vert Q\vert}{M}\leq\frac{\vert\sum\limits_iq_i\vert}{\sum\limits_i m_{g}}\leq\frac{\sum\limits_i\vert q_i\vert}{\sum\limits_i m_{g}}\leq  \max_i \frac{\vert q_i\vert}{m_{i}}, 
\label{ineq}
\end{equation}
 describing  a decay   matching with the following conservation laws
\begin{equation}
M=m_g+\Delta E,\hspace{6pt}Q=\sum_i q_i
\label{eq3}
\end{equation}
where one has used $m_g=\sum\limits_i^{N_{sp}} m_{i}$.  In the present study, furthermore, we assume that  the particles of mass $m_g$ have an  upper  limit of species of maximal  sizes with maximum possible discrete charges.  A similar   description  has been elaborated  for black holes, being endowed with large numbers of species associated with certain  discreet symmetries \cite{SPE1}.  However, it has been observed that it adapts better to a particle of mass $m_g$, by considering the inequality $\sum\limits_i m_{i}\leq \max_{i}  m_{i} = N_{sp}M_p$.
Handling  Eq.(\ref{eq3}) and (\ref{ineq}) in the proposed characteristic function Eq.(\ref{eq2}), we obtain  the following inequality at some fixed black hole mass $M$
\begin{equation}
\mbox{min }\rho\leq\frac{zM^3_pT_p}{32\pi M^2 m_{g}T}e^{-\frac{M^3_pT_p}{32\pi\sum\limits_i m_{i}T}\frac{(\sum\limits_i z_i)^2}{(\sum\limits_i m_{i})^2}}\leq\frac{zM^3_pT_p}{32\pi M^2 m_{g}T}e^{-\frac{M^3_pT_p}{32\pi M^2T}\frac{(\sum \limits_i z_i)^2}{\sum \limits_i m_{i}}},
\end{equation}
where one has used a minimal value of the characteristic function
 \begin{equation}\mbox{min }\rho=\frac{zM^3_pT_p}{32\pi M^2 m_{g}T}\exp\left(-\frac{M^3_pT_p}{32\pi\sum\limits_i m_{i}T} \left( {\max_i} \left(\frac{z_i}{m_{i}}\right)\right)^2\right).  \end{equation} 
 In this way,  the fastest decay channel is described by species  maximizing  the ratio $z_i/m_{i}$, in accordance with  WGC.
 Exploiting,  furthermore,  the bound on $\sum\limits_i m_{i}$,  we get an upper bound on the minimal value of the characteristic function
  \begin{equation}\mbox{min }\rho\leq\frac{zM^3_pT_p}{32\pi M^2 m_{g}T}\exp\left(-\frac{M^2_pT_p}{32\pi N_{sp}T} \left( {\max_i} \left(\frac{z_i}{m_{i}}\right)\right)^2\right).  \end{equation}
In this bound on the minimal value, a cutoff term $\frac{M_p^2}{N_{sp}}$ has appeared. This  allows  one  to confront the proposed instability  behaviors with the  previous works \cite{SPE1,SPE2,SPE3,SPE4,SPE5,SPE6,SPE7,SPE8}. The implications of such cutoffs will be elaborated in the forthcoming   discussions.
\subsection{UV divergence and cutoffs}
\begin{figure}[h]
\begin{center}
\centering
\includegraphics[scale=0.6]{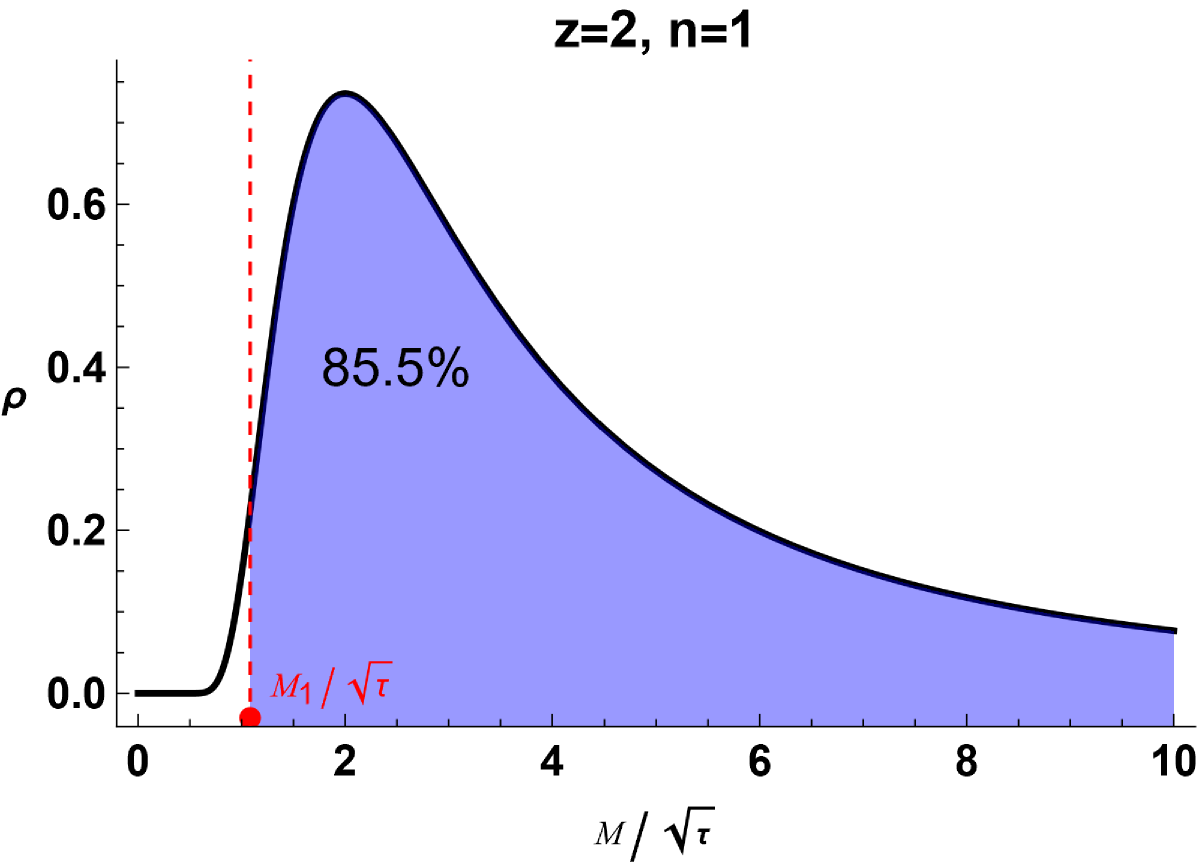}
\caption{Cutoff mas $M_1$ at a degree $n=1$ in the $z=2$ case.}
\label{Fig1}
\end{center}
\end{figure}

In this part,  we discuss some of the ambiguities surrounding the above   cutoff points, in conjunction with WGC.  To do so, we  first reveal 
 the UV divergence behavior  in the proposed characteristic function. This divergence  is encountered by summing over all the possible states in the phase space $\lbrace A,p_\phi,p_\theta\rbrace$. This sum is an integral determined by the cumulative distribution function on the set of random variables $(A,p_\phi,p_\theta)$.  The scenario,  in which the black hole evaporates completely, is a possible interpretation to the convergence of such a function. In this context, the cumulative distribution function  can  be expressed as follows 
\begin{equation}
\mathbb{P}=\iint\rho\frac{dAd^2p}{\hbar^2}=\iint\frac{z}{12k_BTm_gA}e^{-\frac{p^2}{2m_gk_BT}}dAd^2p.
\end{equation}
This integral reduces to 
\begin{equation}
\mathbb{P}=\int^{r_h}_{r_0}\frac{dr}{r} 
\end{equation}
where  $r_h$ is  the even horizon radius of the involved black hole.   It turns out  that  this  integral diverges  by sending  $r_0$  to zero.  To overcome  such  an issue,   the sum needs a cut-off when the probability of accessing states with $r_{0}=0$ is zero. This argument could  be  motivated  by the fact that  the issue  stems  from the infinite sum over inaccessible states.  To explore this issue further, we can treat the cutoff as an interval estimation problem. This could be done by  considering  the characteristic function as a non normalized gamma distribution  given by 
\begin{equation}
\rho=X^{\alpha-1}e^{-\beta X},
\end{equation}
where $\alpha$ and $\beta$ are the shape  and the scale parameters,  respectively.  $X$ is the random variable.
 In the present  investigation $X$  is identified with  $\frac{z\tau}{M^2}$  where one has used   $\tau=\frac{M_p^3T_p}{m_gT}$. In order to avoid small amplitudes, we introduce the normalizing factor $\frac{\beta^\alpha}{\Gamma(\alpha)}$, to be useful. Taking  $\alpha=2$ and $\beta=z$, we can 
calculate the variance of such  a distribution.  Precisely,  we find
\begin{eqnarray} 
Var[\frac{z\tau}{M^2}]=\frac{2}{z^{2}}.
\end{eqnarray} 
In this way, the mass can be treated as a step indexed by an integer. Thus it is expressed in terms of the standard deviation $\sqrt{Var[\frac{z\tau}{M^2}]}$  via  the mean $\langle\frac{z\tau}{M^2}\rangle$ as follows
\begin{eqnarray}
M_n &=& \frac{\sqrt{z\tau}}{\sqrt{\langle\frac{z\tau}{M^2}\rangle+n\sqrt{Var[\frac{z\tau}{M^2}]}}}=  \frac{z\sqrt{\tau}}{\sqrt{2+n\sqrt{2}}}
\end{eqnarray}
where $n$ is an integer indicating the degree of deviation from the mean. Accordingly, the amplitude of accessing a mass value in the rang $[0,\frac{z\tau}{M^2}]$ is  $\frac{\gamma(2,2+n\sqrt{2})}{\Gamma(2)}$  \footnote{ via the computation $\frac{z^2}{\Gamma(2)}\int_0^{\frac{2+n\sqrt{2}}{z}}Xe^{-zX}dX=\frac{\gamma(2,2+n\sqrt{2})}{\Gamma(2)}$.}
where $\gamma(\alpha,x)$ is the lower incomplete gamma function given by the integral $\int^x_0x^{\alpha-1}e^{-x}dx$ \cite{Book3}. As a simple illustration,  Fig.(\ref{Fig1}) shows  the selected sample by considering a degree of deviation $n=1$. Precisely, with such a degree of deviations, one can access a state in the range $[\frac{M_1}{\sqrt{\tau}},\infty[$ with a probability of $85.5\%$.\\

Additionally, by taking  $n=20$,  the amplitude reaches   the value of $99.999\hspace{2pt}999\hspace{2pt}9997797\%$ for $M\geq\frac{z\sqrt{\tau}}{\sqrt{2+20\sqrt{2}}}$.% This means that there is a one in a four hundred billion chance to access a state with $M<\frac{z\sqrt{\tau}}{\sqrt{2+20\sqrt{2}}}$.
   Translating this for normal distributions, we conclude  that  the states bellow a cutoff $M_{20}=\frac{z\sqrt{\tau}}{\sqrt{2+20\sqrt{2}}}$ are inaccessible, with a $7\sigma$ confidence.\\
In the case of many species, Eq.(\ref{eq3}) and  Eq.(\ref{ineq}) describe a mass scale cutoff at $\frac{\sum\limits_i z_i}{\sqrt{2+n\sqrt{2}}}\sqrt{\frac{M_p^3T_p}{32\pi\sum\limits_i m_{i}T}}$. Moreover, considering  such a  upper bound on the mass $\sum\limits_i m_{i}$  the defined cutoff in the proposed  scenario  should satisfy
\begin{equation}
 \frac{M_p}{\sqrt{N_{sp}}}\sqrt{\frac{T_p}{32\pi T(2+n\sqrt{2})}}(\sum\limits_i z_i)<M_{n}.
 \label{eq4}
 \end{equation}
 
% A simple case study, at a step $n=20$ and $(m_g=\frac{M_p}{10},N_{sp}=12)$, The corresponding step for the lower bound on the cutoff is $k=1$, where this %number decrease with the increase of the number of species. Thus, the lower bound on $M_0$ is much more inaccessible then a mass $M=\frac{\sum\limits_i %z_i}{\sqrt{2+n\sqrt{2}}}\sqrt{\frac{M_p^3T_p}{\sum\limits_i m_{g_i}T}}$}.

One can choose $N_{sp}$ to be large enough so that the term $2+n\sqrt{2}$ is ignored. According to \cite{SPE1}, this implies that one has $\Lambda_{sp}\sqrt{\frac{T_p}{32\pi T}}(\sum\limits_i z_i)<M_{n}$, where $\Lambda_{sp}$ represents  the species scale. This observation is consistent with the inverse relation between the species scale $\Lambda_{sp}$ and the number of species $N_{sp}$, which could be linked to the gauge coupling via $N_{sp}\sim \frac{1}{g^2}$, restoring the condition given in Eq.(\ref{WGC}). However, a correction term $\sqrt{\frac{T_p}{32\pi T}}(\sum\limits_i z_i)$ has appeared.\par
Now, we check the consistency of  WGC  with the assumptions that  the black holes with weak gauge couplings are the most common. To do so,  we compute the probability of accessing states   with  either  $M>\frac{Q\sqrt{\tau}}{g}$ or $M<\frac{Q\sqrt{\tau}}{g}$, defined by $\mathcal{P}=\frac{1}{\langle M\rangle}\int^a_b\frac{z\tau}{M^2}e^{-\frac{z\tau}{M^2}}dM$. In order to get such probabilities, we should  integrate along the interval  $]\frac{Q\sqrt{\tau}}{g} ,\infty[$. Taking the black hole mass as the measure, the probability of accessing states with $M>\frac{Q\sqrt{\tau}}{g}$  is found to be 
\begin{equation}
\mathcal{P}(M>\frac{Q\sqrt{\tau}}{g})=erf(1). 
\label{eq5}
\end{equation}
However,  the probability of accessing states with $Q\sqrt{\tau}>M$ is
\begin{equation}
\mathcal{P}(M<\frac{Q\sqrt{\tau}}{g})=cerf(1) 
\label{eq6}
\end{equation}
where  one has used   $erf(1)=\frac{2}{\sqrt{\pi}}\int^1_0e^{-x^2}dx$ and $cerf(x)=1-erf(x)$\cite{Book3}.
It follows from Eq.(\ref{eq5}) and Eq.(\ref{eq6}) that  one has 
\begin{equation}
\mathcal{P}(M>\frac{Q\sqrt{\tau}}{g})>\mathcal{P}(M<\frac{Q\sqrt{\tau}}{g}).
\end{equation}
 This  shows that WGC is consistent with the aforementioned assumptions.   In connection  with the  previous arguments about cutoffs, the probability $\mathcal{P}(M<\frac{Q\sqrt{\tau}}{g})$ decreases by sending  a cutoff $M^\star$  to $\frac{Q\sqrt{\tau}}{g}$. This can  be  supported by  the relation 
\begin{equation}
 \mathcal{P}(M<\frac{Q\sqrt{\tau}}{g})=erf(\frac{Q\sqrt{\tau}}{gM^\star})-erf(1),
 \end{equation}
matching with  Eq.(\ref{eq4}).

\subsection{ Remnant production rate and WGC}
At this stage, we address the implication of the  developed characteristic function in  the remnant production rate.  In black hole physics, the probability of entering a $(N_R,M,g,m_g)$ state is determined as a function of the rate of remnant production and the characteristic function associated with  the asymmetric instability.  Explicitly, such a  probability  is found to be 
\begin{equation}
P(P_{BH}\cup\rho)=\frac{zM^3_pT_p}{32\pi M^2m_gT}e^{2\pi r M_p-\frac{M^2}{M_p^2}-\frac{M^3_pT_pz^2}{32\pi M^2m_gT}}.
\label{eq6}
\end{equation}
At Planck scales, the probability  reduces to  the following form 
\begin{equation}
P(P_{BH}\cup\rho)\sim ze^{-z^2}.
\end{equation}
Thus, it is clear that the asymmetric  instability suppresses the remnant  production rate.

    The suppression of accessible states  could be interpreted for  small scale regimes. As the black hole evaporates down to a small scale, the  horizon  size   decreases.  Due to the Pauli exclusion principle, the pair produced charges can no longer exist on the black hole surface. However,  the existence of  such objects   can be supported by a   disintegrating scenario  into other  stable black holes and particles matching   with  EFTs  in question. This   should follow possible decay paths  with different probabilities. For a black hole evaporating down to a Planck scale,  the only disintegration path is to decay into stable particles as stated by  WGC \cite{INTWGC1,R12}.Lastly, the density function introduced should contribute to the Bekenstein-Hawking entropy, as shown in equation (\ref{eq6}). This leads to significant results concerning the Bekenstein bound, where states are counted over a cubical box. However, the implications of this work for entropy will be the subject of future investigations.

\section{ Swampland statistics   for  charged  black holes}
As applications,  the  present investigation  automatically motivates the discussion  of a charged  solution known by   the Reissner–Nordström black hole in four dimensions \cite{Book4}.   
This black hole is described by the  following metric  line 
\begin{equation}
ds^2=f(r)dt^2-\frac{1}{f(r)}dr^2-r^2(d\theta^2+ \sin^2\theta d\phi^2).
\end{equation}
For this solution,  one has considered
\begin{equation}
f(r)=1-\frac{2GM}{c^2r}+\frac{Q^2G}{4\pi\epsilon_0c^4r^2}
\end{equation}
where  $\epsilon_0$ is the  vacuum  electric permeability.  According to \cite{Book4}, 
the event horizon  can be obtained by solving the constraint  $f(r_h)=0$.   Indeed,  two solutions  can be derived 
\begin{equation}
r_h^{\pm}=\frac{G}{c^2}\left(M\pm\sqrt{M^2-\frac{Q^2}{G\pi\epsilon_0}}\right).
\end{equation}
 According to previous discussions,  the black hole  could involve,   either  a net charge over time due to  the Hawking radiation (a charge  from its formation)  or  both  scenarios.  For a generic situation, the  asymmetric probability densities are  found to be 
\begin{eqnarray}
\rho^+ &=& \frac{z\tau}{M^2(1+\sqrt{1-\frac{Q^2}{G\pi\epsilon_0M^2}})^2}e^{-\frac{\tau z^2}{M^2(1
+\sqrt{1-\frac{Q^2}{G\pi\epsilon_0M^2}})^2}}\\
\rho^- &=& \frac{z\tau}{M^2(1-\sqrt{1-\frac{Q^2}{G\pi\epsilon_0M^2}})^2}e^{-\frac{\tau z^2}{M^2(1
-\sqrt{1-\frac{Q^2}{G\pi\epsilon_0M^2}})^2}}.
\end{eqnarray}
In our case, we only focus  on  $\rho^+$, while a similar analysis  is possible for 
$\rho^-$.  Concretely, the inequality (\ref{ineq1}) becomes
\begin{equation}
\frac{Q}{M}>\frac{g}{\sqrt{{\tau}}}\left(1+\sqrt{1-\frac{Q^2}{G\pi\epsilon_0M^2}}\right).
\label{eq7}
\end{equation}
For real behaviors of the inequality, one should have $\frac{Q}{M}<G\pi\epsilon_0$. This is insured by the inequality
\begin{equation}
\frac{g^2}{\tau}< G\pi\epsilon_0.
\label{eq8}
\end{equation}

For more clarifications the inequality (\ref{eq7}) can be simplified to
\begin{equation}
\frac{Q}{M}>\lambda=\frac{2\frac{g}{\sqrt{\tau}}}{1+\frac{g^2}{\tau G\pi\epsilon_0}}.
\end{equation}

The condition $\sqrt{G\pi\epsilon_0}>\lambda$ is thus insured by the in inequality (\ref{eq8}). It  is worth noting that in terms of charged particle mass,  at $T=T_p$,  the inequality (\ref{eq8})  generates the upper bound 
\begin{equation}
m_g< \frac{M_p}{168\pi\alpha}
\end{equation}
with $\alpha$ the fine structure constant. This bound has been observed to be adhered to

 by the standard model. Therefore, the following Table.(\ref{T1}) contain the different black hole horizon scenarios.

\begin{table}[!h]
\begin{center}
\begin{tabular}{|c|c|c|c|}
\hline
   Black hole Horizon &  Stable   &  Unstable  \\
\hline
   Regular horizon    &  $\sqrt{G\pi\epsilon_0}>\lambda>\frac{Q}{M}$ & $\sqrt{G\pi\epsilon_0}>\frac{Q}{M}>\lambda$ \\
\hline
   Extremal horizon  &   Contradicts inequality (\ref{eq8}) & $\sqrt{G\pi\epsilon_0}=\frac{Q}{M}>\lambda$\\
\hline
   Naked singularity &  Contradicts inequality (\ref{eq8}) & $ \frac{Q}{M}>\sqrt{G\pi\epsilon_0}>\lambda$ \\
\hline
\end{tabular}
\caption{ The main stability scenarios for different black hole horizons}
\label{T1}

\end{center}

\end{table}
%\newpage

 This means that the black hole should disintegrate due to asymmetric instability before reaching its extreme limit, for which no stable solution is possible at this level, as shown in Table (\ref{T1}). Thus, the  naked singularity scenarios can be avoided in accordance with the WGC arguments \cite{INTWGC1}.  Indeed,   we  discern between two limits. One corresponds  to a situation where there is no room to fit more charged states. The second  one asserts that the black hole solution exhibits a naked singularity behavior. Both situations are governed by a different charge to  the mass ratio inequalities.  These assertions would be easily verified in the case of extremal  black holes. Evidently, in order to be unstable, such black holes should satisfy
\begin{equation}
\frac{Q}{M}>\frac{g}{\sqrt{{\tau}}}.
\end{equation}
Considering the fact   that  $\frac{Q}{M}=\sqrt{G\pi\epsilon_0}>\frac{g}{\sqrt{{\tau}}}$, we obtain  a consistent result with Eq.(\ref{eq8}).

%We would like to add, that in order to maximize the charge to mass ratio of the particles/species they must be light species. This fact is well established by the inequality (\ref{eq8}) where light mass imply the following
%\begin{eqnarray}
%\frac{G^2\pi^2\epsilon_0^2M_p^3}{4g^2}<<\mathcal{O}(1),\\
%G<<g.
%\end{eqnarray}
%This, showcase the fact that Gravitational interactions are the weakest in terms of gauge and gravitational couplings

Regarding the remnant production rate, in this scenario, we assume that the black hole evaporates down to a Planck scale by keeping a regular horizon. Considering the  Reissner–Nordström black holes at such small scales, we observe  that the probability of accessing a state $(N_R,M,Q,g,m_g)$   can take the following  form
\begin{equation}
P(P_{BH}\cup\rho_+)\sim\rho_+.
\end{equation}
Consequently, the asymmetric instabilities in these black hole solutions  allow one  to reach the desired result being the suppression of the remnant production rate.

\section{Conclusion}
In this paper, we  have  approached  certain  black hole  issues including  remnant problems by  providing  a statistical description   based on  WGC  in the swampland program.   It has been suggested that  this conjecture basically postulates that,  at  small scales,  the black holes should be unstable.  In fact,  there are possible  decay channels induced by  set instabilities. Based on such arguments,  we  have carried out a statistical investigation  for  non-supersymmetric four dimensional black holes, where such instabilities may be underlined by the Pauli exclusion principle.  Precisely, we  have shown   that this statistical framework allows one  to derive the charge to  the mass ratio inequality  $\frac{Q^2}{M^2}>\frac{32\pi g^2m_gT}{T_pM^3_p}$,  where  $m_g$ is the mass of a particle with  a gauge coupling $g$ and $T$ is a thermodynamical equilibrium temperature with Planck scales. Consequently,  it has been remarked that this formulation is similar to that of the weak gravity  swampland conjecture, as the proposed instabilities also appear at small scales. Supported by the proposed instabilities, we  have provided  a suppression mechanism to the remnant  production rate.  Moreover, we  have   considered a generic situation  to address the species scale. Specifically, we have treated the cuts in the mass scale of the black hole $M_0$ as an interval estimation problem.  For a  generic situation,  this  has resulted  in a black hole mass cutoff lower bound $ \frac{M_p}{\sqrt{N_{sp}}}\sqrt{\frac{T_p}{32\pi T}}(\sum\limits_i z_i)<M_n$, where $z_i$ and $N_{sp}$  are the number of charged states and the number species, respectively.  We have observed   that the upper bound  can be considered as a  corrected form of the species scale introduced in the literature. Using the developed formalism of the proposed instabilities, we  have reconsidered the study of the Reissner–Nordström black holes as a   possible   application for charged solutions. As a  result, we have found that this  illustrating model is consistent with the previous observations, including the  remnant production rate suppression. Furthermore, we have revealed that  the proposed instabilities prohibit naked singularity  behaviors   due to the  constraint  $\frac{g^2}{\tau}< G\pi\epsilon_0$.

This work   leaves certain open issues.   A natural question concerns extra black hole backgrounds including  either the rotating parameter or  extra  scalar fields  originated  from different scenarios. A possible road is to implement  stringy scalars which could be derived from  superstring  compactification mechanisms via  internal  compact spaces with non-trivial holonomy groups  such as  Calabi-Yau geometries.

\section*{Conflicts of interest}
The authors declare that they have no conflicts of interest.

\end{document}